 \documentclass[10pt, conference, letterpaper]{IEEEtran}
\usepackage{cite}

\ifCLASSINFOpdf
  \usepackage[pdftex]{graphicx}
  \graphicspath{{../pdf/}{../jpeg/}}
  \DeclareGraphicsExtensions{.pdf,.jpeg,.png}
\else
   \usepackage[dvips]{graphicx}
   \graphicspath{{../eps/}}
   \DeclareGraphicsExtensions{.eps}
\fi
\usepackage{amsmath}
\usepackage{array}

\ifCLASSOPTIONcompsoc
  \usepackage[caption=false,font=normalsize,labelfont=sf,textfont=sf]{subfig}
\else
  \usepackage[caption=false,font=footnotesize]{subfig}
\fi
\begin{document}
%
\title{BFR: a Bloom Filter-based Routing Approach for Information-Centric Networks}

\author{\IEEEauthorblockN{Ali~Marandi\IEEEauthorrefmark{1},
Torsten~Braun\IEEEauthorrefmark{1}, Kav\'e Salamatian\IEEEauthorrefmark{2} and
Nikolaos~Thomos\IEEEauthorrefmark{3}}
\IEEEauthorblockA {
 \IEEEauthorrefmark{1}University of Bern, Bern, Switzerland \\ Email:\{marandi,braun@inf.unibe.ch\}\\
 \IEEEauthorrefmark{2}Universit\'e de Savoie, France\\Email: kave.salamatian@univ-savoie.fr\\
 \IEEEauthorrefmark{3}University of Essex, Colchester, United Kingdom\\ Email: nthomos@essex.ac.uk
}}


\maketitle

\begin{abstract}
Locating the demanded content is one of the major challenges in Information-Centric Networking (ICN). This process is known as content discovery. To facilitate content discovery, in this paper we focus on Named Data Networking (NDN) and propose a novel routing scheme for content discovery, called Bloom Filter-based Routing (BFR), which is fully distributed, content oriented, and topology agnostic at the intra-domain level. In BFR, origin servers advertise their content objects using Bloom filters. We compare the performance of the proposed BFR with flooding and shortest path content discovery approaches. BFR outperforms its counterparts in terms of the average round-trip delay, while it is shown to be very robust to false positive reports from Bloom filters. Also, BFR is much more robust than shortest path routing to topology changes. BFR strongly outperforms flooding and performs almost equal with shortest path routing with respect to the normalized communication costs for data retrieval and total communication overhead for forwarding Interests. All the three approaches achieve similar mean hit distance. The signalling overhead for content advertisement in BFR is much lower than the signalling overhead for calculating shortest paths in the shortest path approach. Finally, BFR requires small storage overhead for maintaining content advertisements.

\end{abstract}



%
\IEEEpeerreviewmaketitle

\section{Introduction}
\label{sec::intro} 
NDN is one of the most prominent ICN \cite{ICNsurvey} proposals that aim to enhance and/or replace the current IP-based communication model. It is based on hierarchical names for content objects, in-network caching mechanisms, and content-level security. NDN pursues at first a long time goal of the networking community, {\em{i.e.}}, providing location independence to communication \cite{locationindependence}. To reach this goal, content retrieval should be content-oriented, decoupling content objects from locations so that users can retrieve them even if their locations change. There are two packet types in NDN: \emph{Interest} and \emph{Data}. Users issue requests for content sending Interests, which carry hierarchical content names rather than IP addresses. Therefore, lookup and routing operations are based on hierarchical content names as well. 

Each NDN node uses three main data structures: \emph{Content Store (CS)}, in which each node caches the received content, \emph{Pending Interest Table (PIT)}, where nodes maintain received Interests and the faces through which the nodes receive them, and the \emph{Forwarding Information Base (FIB)}, in which each node maintains information about the next hop face(s) through which known name prefixes can be reached. 
In NDN, routing operations are performed only for Interests, meaning that Data packets use the traces left by Interests in the corresponding PIT entries, and follow the reverse path of Interests to determine the locations of the content requesters. 
An Interest packet has a \emph{nonce} field, which contains a random value. This field is used to detect and discard duplicate Interests coming from different paths. Hence, loop freedom is ensured for Interests. 

In NDN, users issue Interests to request Data packets. It is necessary to route each Interest over the path(s) through which it can reach the demanded Data. Hence, routing on content names is a very important problem in ICN. To route an Interest, each node looks up the name of the Interest performing a Longest Prefix Matching (LPM) operation in the FIB. If there is a FIB entry that contains information about  the next hop face(s) for the name of the Interest, or a prefix of it, the Interest will be forwarded through the next hop face(s) that are recorded in the corresponding FIB entry. Therefore, the development of strategies that optimally populate FIBs is vital for NDN. This has been the focus of many proposed routing protocols  \cite{nlsr, advertising, scan, cobra}.  When FIBs are populated, the forwarding {\emph{strategy}} decides the face(s) over which an Interest should be forwarded from among the next hop face(s) specified in the matching FIB entry for that Interest.  For example, the {\emph{multicast}} strategy forwards an Interest over all the faces specified in the matching FIB entry.

In general, there are two main classes of content discovery solutions in ICN, namely: resolution-based and routing-based. Resolution-based solutions map requesters with content producers at rendezvous points \cite{netinf, mdht, pursuit}. These schemes have small traffic footprints, but their performance degrades when there is large and dynamic content demands. Routing strategies \cite{breadcrumbs, nlsr, advertising, inform, scan, dht, cobra}, such as {\em{Flooding}} or algorithms based on {\em{Shortest path}} calculations, explore a larger area of the network than resolution-based solutions, and hence, have a higher chance of finding the content \cite{scopedflooding}. The \emph{Flooding} method forwards all the Interests through all the faces except the incoming one. This makes flooding inefficient as it wastes significant bandwidth resources. Differently from flooding, \emph{Shortest path} routing solutions forward each Interest only over the shortest path to the origin server of the demanded content object. These routing solutions require full knowledge of the topology as well as the location of origin servers for all the existing name prefixes in the network that entails a large overhead.
To avoid wasting network resources through Interests flooding, an alternative approach is to permit origin servers advertising their content offers frequently, {\em{i.e.}}, whenever new content objects are available in repositories. Therefore, origin servers could represent their content offers using Bloom Filters (BFs) that can represent sets in a compact way. This leads to a smaller overhead needed for the propagation of content advertisements. Due to these appealing features of BF-based content advertisement, in this paper we propose BFR, a routing protocol that uses BFs for content advertisements from origin servers for FIB population.

In NDN, temporary copies of a content object might be cached en-route to the nodes that provide the permanent copies of the content object. This possibility of in-network caching enables consumers to retrieve content objects from the caches that are closer than servers. In our scheme only origin servers perform BF-based content advertisement. Nevertheless, nodes receive the content advertisement of an origin server from all the paths en-route to the origin server and populate their FIBs accordingly. Further, we adopt the multicast strategy for forwarding Interests. Therefore, BFR forwards each Interest in parallel through all the paths towards the origin server of its demanded content object. The Interest could be satisfied from the caches before reaching the origin server. Hence, it is unnecessary for routers to explicitly advertise their cached content objects, like the scheme proposed in \cite{prefetching}, and incur more advertisement overhead. 




BFR is topology oblivious. Hence, it does not need to propagate and store information about the topology that entails overhead. In addition, BFR requires reasonable storage and signalling overhead for content advertisements. Further, it does not adopt any IP-based routing protocol as primary or fall-back mechanism.  This makes BFR fully content oriented, and removes any dependencies on IP-based communication models. 

The remainder of this paper is organized as follows. We discuss related work in Section \ref{sec::rw}. Section \ref{sec::sd} describes the proposed BFR method. 
Then, Section \ref{sec::dis} discusses the impact of false positive errors on BFR operation, robustness to topology changes, and handling of content migration.
Afterwards, we present in Section \ref{sec::pe} a simulation-based comparative analysis of the proposed BFR against flooding and shortest path schemes to illustrate BFR advantages in practice.  Section \ref{sec::con} concludes the paper.

\section{Related work}
\label{sec::rw}
Content discovery using routing in Information-Centric Networks has been previously proposed in \cite{breadcrumbs, nlsr, advertising, inform, scan, dht, cobra }. In \cite{breadcrumbs}, nodes store so-called \emph{breadcrumbs}, {\em{i.e.}}, the traces left from already retrieved content objects along the downstream path towards content requesters to perform routing. Thus, breadcrumbs can be used by routers to route repetitive Interests. Inspired by \cite{breadcrumbs}, in \cite{cobra}, it is proposed to use one Stable Bloom Filter (SBF) per face to record the traces of Data packets passed through each face. SBF is a Counting Bloom Filter (CBF). CBFs are represented by an array of $n-bit$ counters rather than a bit table. The advantage of CBFs over BFs is that deleting an element is allowed by decreasing the counters associated to it, while it is not possible to delete elements from a BF. In SBF, each counter is composed of $d$ bits. When an element is inserted, $P$ counters are randomly selected, and their values are decreased. Further, the $K$ cells associated with the inserted element are set to the maximum value, $2^d -1$. Taking these two actions in parallel keeps the proportion of $0$'s and $1$'s constant, and automatically removes the stale content objects from the SBF. The schemes in \cite{cobra} make use of this property in order to maintain only the traces for effective content objects that are still retrievable through a face in the SBF associated to it. The disadvantage of the approaches in \cite{breadcrumbs} and \cite{cobra} is that they end up in flooding the Interests issued for the first time in the network because there are no stored breadcrumbs for them.  

  In \cite{prefetching}, it has been proposed to collect in a BF all the content objects resolvable by each router to avoid \emph{flooding} of Interests. This approach is similar to the well-known ``summary cache'' scheme \cite{summarycache}, in which BFs are exchanged between web caches as content summaries. This method, however, still does not avoid flooding, since each Interest issued for the first time in the network has to reach the nodes that have permanent copies of the demanded content object and cannot retrieve the demanded content object from temporary caches. Therefore, spreading information about content objects cached at routers does not prevent flooding the network completely. 
 Furthermore, the approach presented in \cite{prefetching} has not been implemented and evaluated. In  \cite{advertising, scan, dht}, BFs are used in order to compress FIBs.  

In \cite{scan}, SCAN is proposed as a routing scheme for content-aware networks. The main disadvantage of SCAN is that it uses IP routing as a fall-back solution, meaning that cache routers perform both content and IP routing, {\em{i.e.}}, they maintain content routing tables as well as IP routing tables. Therefore, SCAN is not a fully content-oriented routing scheme. 

NLSR \cite{nlsr} is considered as one of the most prominent routing-based solutions for NDN. It is a link state routing protocol, which requires frequent pulling of routing updates. NLSR routing updates contain information about both topology and content name prefixes. 
In NLSR, nodes run the Dijkstra algorithm to find the shortest path from each of the faces for any incoming Interest using full information about the topology and the content prefixes that exist in the network. Compared to NLSR, our scheme does not require any knowledge about the topology, while it permits the origin servers to propagate compact content advertisements using BFs.


\section{Bloom filter-based Routing}
\label{sec::sd}
In BFR, origin servers represent and advertise their content objects using BFs.  
In summary, BFR consists of three phases: a) Representation of content objects using BFs, b) BF-based content advertisement, and c) Content retrieval and FIB population. In the following, we describe each phase in detail.

\subsection{Representation of content objects using BFs} 
BF is a space-efficient data structure to represent sets in a compact way and to support membership queries.  When one represents a set with a BF, false positive probability controls the performance of the BF, {\em{i.e.}}, the probability that an element that is not in the set is wrongly reported by a BF as being in the set. In \cite{bloom}, the false positive probability is expressed as a function of the length of bit table $m$, the length of the original set represented by BF $n$, and the number of hash functions $k$. According to \cite{bloom}, when one wants to insert $n$ elements in a BF and can afford a false positive probability $p$, the required size for the bit table $m$ and the number of hash functions $k$ are respectively given as:
\begin{equation}
\label{eq::bf}
 \begin{cases}
   m = -\frac {nln(p)} {(ln2)^2} \\
   k = \frac{m}{n} {ln2}
 \end{cases}
\end{equation}

\begin{figure}[t]
\[
 \begin{cases}
e_1=/unibe.ch/,\\  e_2=/unibe.ch/images/,\\ e_3= /unibe.ch/images/fileName1
\end{cases}
\]
\centering
\includegraphics[width = 0.8 \columnwidth]{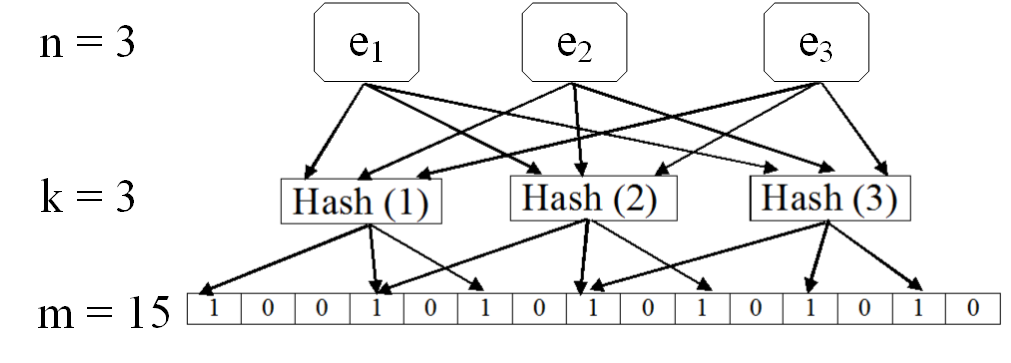}
\caption{An example for content advertisement BF and related hash functions}
\label{fig::BF}
\vspace{-1.1em}
\end{figure}

 In BFR, when an origin server has content objects to offer, it generates an empty BF for which all the $m$ bits of the bit array are set to zero. Then, the origin server maps the names of  its content objects into the generated BF. An example of inserting three URLs into a BF with a parameter set $\{m=15, n=3, k=3\}$  is presented in Fig. \ref{fig::BF}. As Fig. \ref{fig::BF}  shows, the insertion process consists of feeding each URL to the three hash functions to get three positions in the bit array and set all the bits at these positions to $1$. 


In BFR, each origin server maps the names of its content objects as well as their name prefixes in its BF. For example, as Fig. \ref{fig::BF} shows, the full name (e.g., $/unibe.ch/images/fileName1$) as well as the name prefixes (e.g.,$/unibe.ch/$ and $/unibe.ch/images/$) are inserted into the BF. In Section \ref{subsec::fibpop}, we discuss the reasons behind inserting name prefixes into BFs in detail.


%
%

To show the savings resulting from using BFs for representing a set of content objects, we provide an example. Consider that an origin server stores $200$ content objects, which are each divided into a number of segments. To represent the content objects, the server creates a BF by setting $n=200$, and targets a false positive error probability of $2\%$ (approximately four names per BF). Thus, the server needs a BF of size $m=1628.47$ bits and $k=5.64$ hash functions. Aligning the bit table size to byte order and rounding these values, the server requires $203.5$ bytes, {\em{i.e.}}, approximately one byte per named content object. For larger BFs, {\em{i.e.}}, larger values of $m$ and $n$, and the same false positive probability, the required space for inserting each URL into the BF stays constant, {\em{i.e.}}, one byte. In NDN, names are URLs. To evaluate our routing approach, we consider a realistic URL catalogue \cite{selfsimilarity} with the average URL size equal to $42.45$ bytes. For this setting, a server needs $8490$ bytes to advertise a list of $200$ URLs without BF, while it needs only $2.4\%$ of this size, {\em{i.e.}}, $203.5$ bytes, in case it uses BF. Therefore, the use of BFs  results in high compression for representing a set of content objects.  



\subsection{BF-based content advertisement}
\label{subsec::ca}
When an origin server creates a BF that contains the names of its content objects, it propagates this BF to advertise its content objects. 
To propagate the content advertisement BF and be compatible with the original NDN, an origin server could encapsulate the BF in an Interest or a Data packet. If the content advertisement BFs would be inserted in Data packets, all the nodes but the server nodes, {\em{i.e.}}, routers and consumers, should pull the content advertisement messages. Such a strategy is followed in \cite{nlsr}, where all the nodes frequently pull routing information regarding the topology and name prefix updates from the neighbourhood. However, in BFR only certain nodes, {\em{i.e.}}, origin servers, propagate routing information ({\em{i.e.}}, content advertisements) and the rest of the nodes are unaware of the locations of the origin servers. Therefore, it is not clear up to which scope the content advertisements should be pulled.

\begin{figure}[t]
\centering
\includegraphics[width = 0.85\columnwidth]{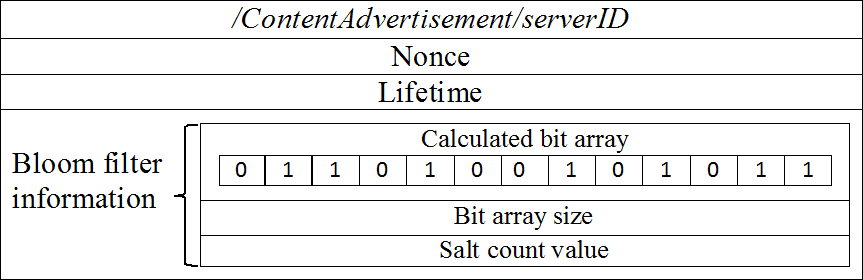}
\caption{CAI message}
\label{fig::CA}
\end{figure}

To address this problem, we opt for a push-based content advertisement scheme. We introduce a new type of Interest packet called Content Advertisement Interest (CAI) that carries content advertisement BFs. Hence, BFR propagates CAI messages to propagate the content advertisement BFs. The NDN Interest forwarding pipeline detects and discards duplicate CAI messages and ensures loop freedom for these messages. It is important to note that the only purpose for the propagation of CAI messages is content advertisement and no Data packet is sent as a response to CAI messages. Fig. \ref{fig::CA} illustrates the structure of a CAI message that is identified by the name prefix $/ContentAdvertisement$. To distinguish the CAI messages issued by different origin servers, we allow each origin server to append its unique ID as the second name component to the name of the CAI messages that it issues. In the forthcoming, we describe the reasons behind this choice in detail. As Fig. \ref{fig::CA} shows, each CAI message similar to Interest messages exploits a random {\em{nonce}} to ensure loop freedom. The nodes that receive CAI messages store them in their PITs. CAI messages should expire like other packet types stored in nodes' PITs. Since no Data is coming back in response to the CAI messages, they stay in PITs until their timeout. Hence, it is necessary to add to the CAI message a {\emph{lifetime}} field, which indicates when it expires. 
To this aim, we reuse the {\emph{Interest lifetime}} field to indicate the lifetime of CAI messages. Origin servers refresh the CAI messages to keep nodes informed about their content offers. Further, the content advertisement applications do not re-express the CAI messages. We should emphasize that this work aims at proposing a BF-based content advertisement strategy fully compatible with the original NDN and not to present a NDN variation. The last components for a CAI message are the needed information to retrieve the content advertisement BF consists of the calculated bit array, the size of the bit array, and a salt count value that is needed to retrieve the same content advertisement BF at the nodes that receive the CAI message. Here, we assume that all the origin servers generate their hash functions with a universal random seed and operate with the same set of hash functions.


To permit nodes to propagate CAI messages, we add a FIB entry for name prefix $/ContentAdvertisement$ in the FIBs of all the nodes, and add all the faces as next hops for this name prefix at each node. Further, we adopt the multicast strategy for forwarding the $/ContentAdvertisement$ name prefix. Therefore, when an origin server issues a CAI message, this message is forwarded to all the nodes that are located in one hop distance and those nodes forward it over all the faces except the incoming one. Each node that receives the CAI message broadcasts it, while the Interest forwarding pipeline of NDN Forwarding Daemon (NFD) ensures loop freedom and discards duplicate CAI messages. Therefore, all the nodes will eventually receive the CAI message.

%

\begin{figure}[t]
\centering
 \includegraphics[width=0.85\columnwidth]{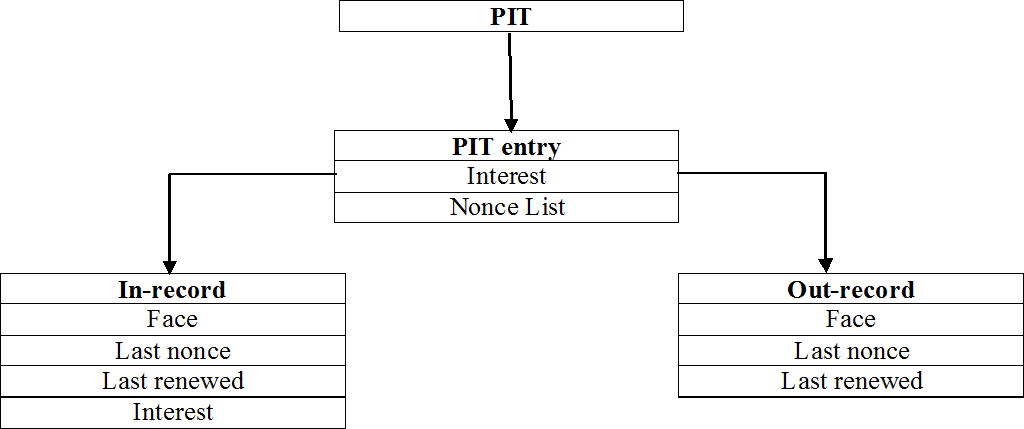}
 \caption{PIT and related entries}
 \label{fig::pitentry}
\vspace{-1em}
\end{figure}

 The nodes that receive CAI messages record in their PITs the faces over which they receive each CAI message. Fig. \ref{fig::pitentry} illustrates the structure of a PIT entry in NDN. As Fig. \ref{fig::pitentry} shows, the faces over which an Interest is received are stored in the {\emph{in-records}} of the related PIT entry. Therefore, to record the faces over which a CAI message is received, we make use of in-records. 

\begin{figure}[t]
  \vspace{-1em}
\centering
 \includegraphics[width=1\columnwidth]{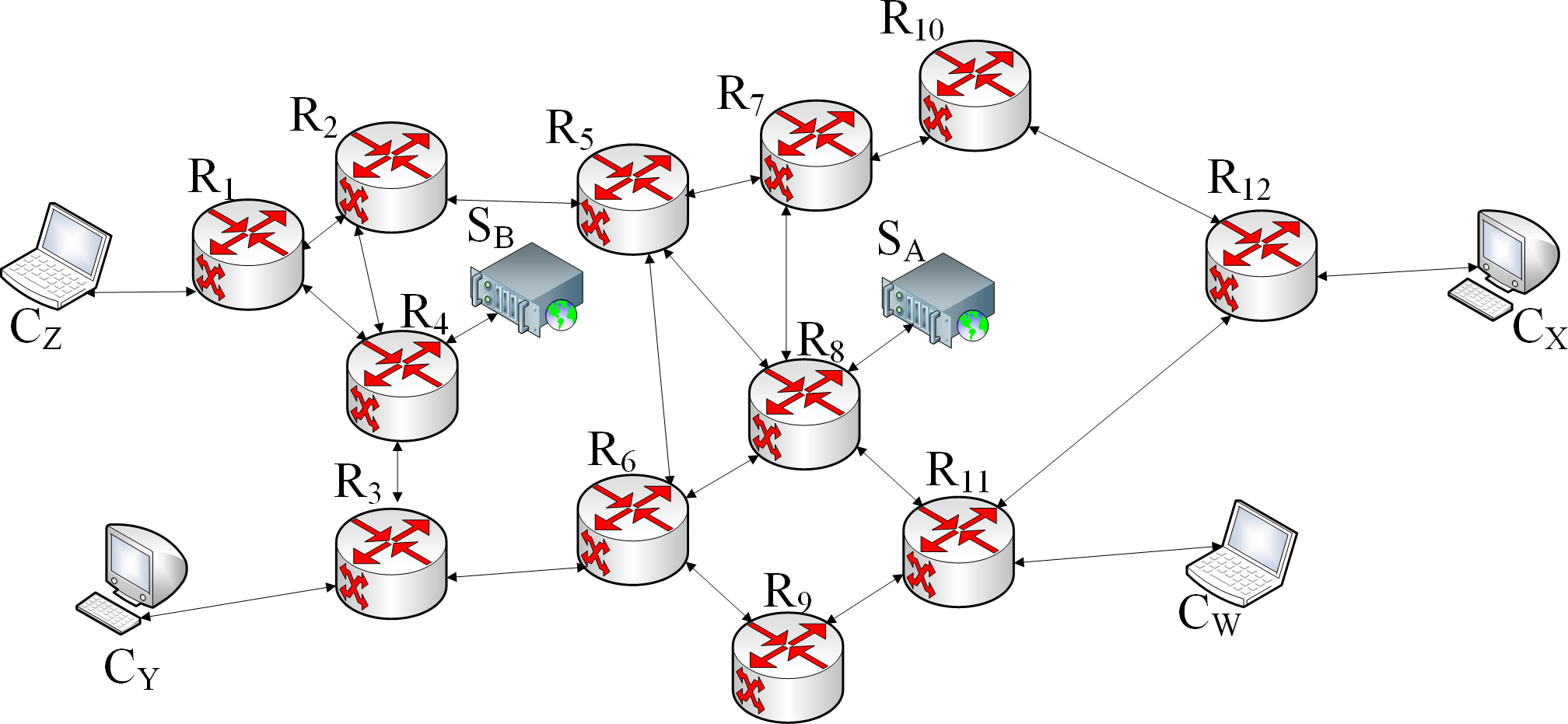}
 \caption{Proposed BF-based content advertisement}
 \label{fig::CAp}
 \vspace{-1em}
\end{figure}

   All the CAI messages, issued by the origin servers, share the same name prefix, {\em{i.e.}}, $/ContentAdvertisement$. Nevertheless, we  let origin servers append their $uniqueIDs$ as second name component to the name of the CAI message. For example, in Fig. \ref{fig::CAp} server $S_A$ generates a CAI message with name prefix $/ContentAdvertisement/A$ and server $S_B$ a CAI message with name prefix $/ContentAdvertisement/B$. In general, servers could append any kind of unique ID ({\em{e.g.}}, their MAC addresses) as the second name hierarchy to ensure name uniqueness. Let us provide an example to explain the reason behind appending $serverIDs$ as the second name hierarchy for CAI messages. In Fig. \ref{fig::CAp}, imagine that servers $S_A$ and $S_B$ do not append their unique IDs as the second name component to CAI messages. In such a case, if server $S_A$ sends a CAI message with name $/ContentAdvertisement$, and router $R_5$ receives it. If at a later time instant, server $S_B$ sends a CAI message with the same name, which is received also by router $R_5$, the Interest forwarding pipeline of NFD will consider the second CAI message received by router $R_5$ as redundant because both messages have the same name. This will lead router $R_5$ to only record the incoming face of the content advertisement issued by server $S_B$ in the PIT entry for name prefix $/ContentAdvertisement$ and to discard it. This approach makes router $R_5$ to discard the content advertisement BF of server $S_B$. Hence, router $R_5$ will be unaware of the content offers from server $S_B$. To avoid this problem, in BFR origin servers append their $uniqueIDs$ to the name of CAI messages.  

To illustrate the content advertisement process,  assume that in Fig. \ref{fig::CAp} server $S_A$ starts the content advertisement process by sending a CAI message to router $R_8$. This router receives and stores this message in its PIT, and forwards it to other nodes, {\em{i.e.}},  routers $R_5$,  $R_6$, $R_7$, and $R_{11}$. Other nodes also store the CAI message in their PITs and forward it over all the faces except the face over which the message has been received. This is done until all the nodes obtain the CAI message. At the end of this process, all the nodes receive the CAI message issued by server $S_A$. In general, CAI messages could flood the network, or could be sent using random walk. Although the random walk strategy incurs less bandwidth and storage overhead, we did not follow this strategy because not all the nodes will be aware of the content objects offered by all the origin servers.

\subsection{Content retrieval and FIB population}
\label{subsec::fibpop}


BFR combines content retrieval and FIB population processes. To begin with the description of the FIB population process, assume that server $S_B$ in Fig. \ref{fig::CAp} also advertises its content offers. After the completion of the content advertisement propagation from servers $S_A$ and $S_B$ at time instant $t_2$, all the nodes store the CAI messages $/ContentAdvertisement/A$ and $/ContentAdvertisement/B$ in their PITs. The PIT of consumer $C_Z$ is presented in Table \ref{tab::PIT}. This Table shows the CAI messages in the upper rows of the PIT to indicate that the CAI messages are distributed proactively. In BFR, nodes use the received CAI messages for FIB population. When a consumer issues an Interest to retrieve some Data, FIB population occurs hop by hop at all the nodes that are placed on paths en-route to the origin server of the demanded Data.


Let us describe FIB population, by considering the topology presented in Fig. \ref{fig::CAp} and assuming that at time $t_3$, consumer $C_Z$ issues the first transmission of the Interest $/unibe.ch/images/fileName1/01$ to retrieve the first segment of content object $/unibe.ch/images/fileName1$ that is offered by server $S_A$. To populate its FIB, consumer $C_Z$ eliminates the sequence number from the name of the issued Interest and checks whether the BFs of the stored CAI messages contain name prefix  $/unibe.ch/images/fileName1$.

\begin {table}[t]
\caption {PIT table of $C_Z$}
 \vspace{-1.0em}
 \label{tab::PIT}
\begin{center}
\begin{tabular}{|c|}
\hline
$/ContentAdvertisement/A$ \\ \hline
$/ContentAdvertisement/B$ \\ \hline
$/unibe.ch/images/fileName1/01$ \\ \hline
\end{tabular}
\end{center}
  \vspace{-1.5em}
\end {table}

 In this case, the demanded content object is produced by server $S_A$, so the BF stored in $/ContentAdvertisement/A$ verifies that it contains the name prefix  $/unibe.ch/images/fileName1$. Now, consumer $C_Z$ can add the face(s) over which it has received content advertisement $/ContentAdvertisement/A$ as the next hop faces for name prefix $/unibe.ch/images/fileName1$ into the FIB.  Therefore,  if no FIB entry exists for this name prefix, consumer $C_Z$ creates a FIB entry for  this name prefix and adds the face(s) stored in the in-records for the CAI message $/ContentAdvertisement/A$ as the next hop face(s) for the FIB entry. After consumer $C_Z$ has populated its FIB for name prefix $/unibe.ch/images/fileName1$, it forwards the Interest for this name prefix to router $R_1$. This router runs the same process as consumer $C_Z$ and checks the Interest name without sequence number, {\em{i.e.}},  $/unibe.ch/images/fileName1$ in the BF of CAI messages stored in its PIT. Router $R_1$ continues to forward the Interest to other nodes according to multicast strategy until the Interest reaches server $S_A$ and the demanded content object is retrieved. The first transmission of Interest $/unibe.ch/images/fileName1/01$ in the network should reach server $S_A$ to retrieve the demanded content object. The next transmissions of this Interest, may retrieve the content object from closer caches at routers situated en-route the upstream path towards server $S_A$.  

We select the \emph{multicast} forwarding strategy that forwards the received Interests over all the next hops specified in the FIB for their names and design BFR to work with this strategy to benefit from the existence of multiple paths between the consumers and the content servers. This approach is very efficient in case of topology changes, {\em{i.e.}}, unexpected links' failures or recoveries or when the shortest paths are congested, thus not able to return Data packets fast enough.

\section{Discussion}
\label{sec::dis}
Here, we discuss the impact of false positive errors on BFR operation, robustness to topology changes, and handling of content migration.

\subsection{Impact of false positive errors on BFR operation}
\label{subsec::falsepositive}
When we use BFs, false negative errors cannot happen, however, false positive errors are possible and affect the performance of the system. To discuss the impact of false positive errors on BFR operation, let us again study the example in Section III-C. Assume that all the content advertisement BFs operate with the false positive probability of $2\%$ and all the content advertisement BFs of an origin server operate with the same set of hash functions. Consumer $C_Z$ issues Interest $I_p$ for name prefix $p = /unibe.ch/images/fileName1$, while server $S_A$ possesses the content objects for this name prefix. When consumer $C_Z$ checks name prefix $p$ in the BFs of CAI messages  $/ContentAdvertisement/A$ and  $/ContentAdvertisement/B$, the BF stored in the former CAI message correctly verifies that it contains name prefix $p$ because false negative errors cannot happen. However, the BF stored in the latter CAI message may falsely report with probability $2\%$ that it contains name prefix $p$. If this false positive report happens, consumer $C_Z$ populates its FIB for name prefix $p$ according to the faces stored in the in-records of both CAI messages $/ContentAdvertisement/A$ and $/ContentAdvertisement/B$. Therefore, consumer $C_Z$ routes the Interest for name prefix $p$ towards both servers $S_A$ and $S_B$. Router $R_1$ receives the Interest for name prefix $p$ from consumer $C_Z$. At this router, the content advertisement BF of CAI message $/ContentAdvertisement/A$ correctly verifies that it contains name prefix $p$ and, therefore, router $R_1$ forwards the Interest for this name prefix  towards server $S_A$. However, at the same router, the content advertisement BF of CAI message $/ContentAdvertisement/B$ might give a false positive report for name prefix $p$ because all the content advertisement BFs of server $S_B$ operate with same hash functions. Therefore, router $R_1$ might forward the Interest for name prefix $p$ towards server $S_B$ as well. When routers $R_2$ and $R_4$  receive the Interest $I_p$ from router $R_1$, they make the same forwarding decisions as router $R_1$. Also, subsequent nodes that receive Interest $I_p$ from routers $R_2$ and $R_4$  take the same forwarding actions. Hence, Interest $I_p$ will be eventually satisfied because all the nodes forward it towards server $S_A$, which provides the demanded content object. However, this Interest might reach server $S_B$, which does not provide the demanded content object.  In summary, when node $n$ checks an Interest for name prefix $p$ against all the content advertisement BFs stored in the PIT, the one that contains name prefix $p$ correctly verifies that it has this name prefix because false negative reports are impossible for BFs. At the same node, if another  content advertisement BF gives a false positive report for name prefix $p$, the Interest will be forwarded towards both the correct origin server, {\em{i.e.}}, the origin server, which provides the demanded content object and the wrong origin server, {\em{i.e.}}, the server that does not provide the demanded content object. This forwarding pattern leads the Interest to be satisfied anyway because it is forwarded over the paths towards the origin server of the demanded content object, while the Interest might reach a wrong origin server due to several wrong forwarding decisions caused by false positive reports from content advertisement BFs at several nodes

\subsection{Robustness to topology changes}
To combat link failures, routing protocols should be resilient to link failures and should adapt to link recoveries. When a link failure is detected, the nodes connected to the failed link should prevent Interests from passing through this link until it recovers. This is done in BFR by taking the following actions, when a node detects a link failure: a) the node removes the face associated with the failed link from all the in-records of all the CAI messages that exist in the PIT, and b) it removes the face associated with the failed link from all the FIB entries 

When detecting a recovered link, the nodes connected to this link force all the Interests to pass through it as it is a newly allocated network resource. In BFR, the nodes connected to a recovered link perform the following actions: a) they add the face associated to the recovered link to all the in-records of all the CAI messages that exist in the PIT, and b) they add the face associated to the recovered link as a next hop face in all the FIB entries. It is worth nothing that the Interests pass through a recovered link for a short time because by receiving fresh CAI messages, all the routes will be automatically updated.





\subsection{Handling of content migration}
\label{sec::cm}
Content migration, {\em{i.e.}}, moving a number of content objects stored in the repository of a server to the repository of another server, may occur in networks. When content migration happens, it is necessary to propagate new CAI messages and to immediately inform the network about the changes in the servers' repositories so that nodes remove the stale CAI messages stored in PITs. For this purpose, we present a strategy, which aims at removing stale CAI messages from PITs upon detecting a content migration event. Let us explain our strategy by considering again the topology illustrated in Fig. \ref{fig::CAp}. Assume that consumer $C_Z$ maintains CAI messages from servers $S_A$ and $S_B$ in the PIT. If server $S_A$ migrates content objects to server $S_B$, these servers immediately propagate new CAI messages in order to inform all the network nodes about this event. However, servers $S_A$ and $S_B$ should not only update the nodes with new CAI messages, but they should also signal them to discard the CAI messages received before. For this reason, we enable servers to do this by adding a new flag called {\emph{discardOldAdverts}} to the new CAI messages. Therefore, servers $S_A$ and $S_B$ activate the {\emph{discardOldAdverts}} flag for the new CAI messages and propagate them. When consumer $C_Z$ receives the new CAI messages in which the {\emph{discardOldAdverts}} flag has been activated, it removes all the CAI messages received in the past, which have been issued by $S_A$ and $S_B$ from its PIT, and stores the new CAI messages. When an origin server replicates content objects to cache servers, cache servers also should advertise their content objects. If the content advertisment BF of a cache server is identical with a content advertisement BF of an origin server, the nodes that receive these identical BFs can aggregate them. Origin servers might add or remove content objects to/from their repositories. If an origin server adds content objects to the repository, it advertises the fresh content objects at the next content advertisement round. If an origin server removes content objects from the repository, the removed content objects will not be inserted in the content advertisement BF next time that the origin server advertises its content objects. In case an origin server receives an Interest for a content object that it has removed recently, the origin server returns a ``No Data" NACK \cite{stateful}  to announce the removal of the demanded content object.

\section{Performance evaluation}
\label{sec::pe}
%
We compare BFR with two other routing approaches:  {\emph{flooding}}, where an incoming Interest is forwarded to all the faces except the incoming one and  {\emph{shortest path}}, where the Dijkstra algorithm is employed to calculate the shortest paths, in terms of least number of hops to the origin servers. We do not disable in-network caching for any of the compared routing strategies. Also, we adopt the multicast forwarding strategy for BFR. We implemented the proposed BFR as well as flooding and shortest path routing strategies in the ndnSIM2.1 \cite{ndnsim2.1} environment. 
We introduced in ndnSIM2.1 some ad-hoc functionalities to reproduce the behaviour of BFR, {\em{i.e.}}, BF-based content advertisement, BF-based FIB population, etc.

\subsection{Simulation settings}
\begin{figure}[t]
  \vspace{-0.5em}
\centering
 \includegraphics[width=0.8\columnwidth]{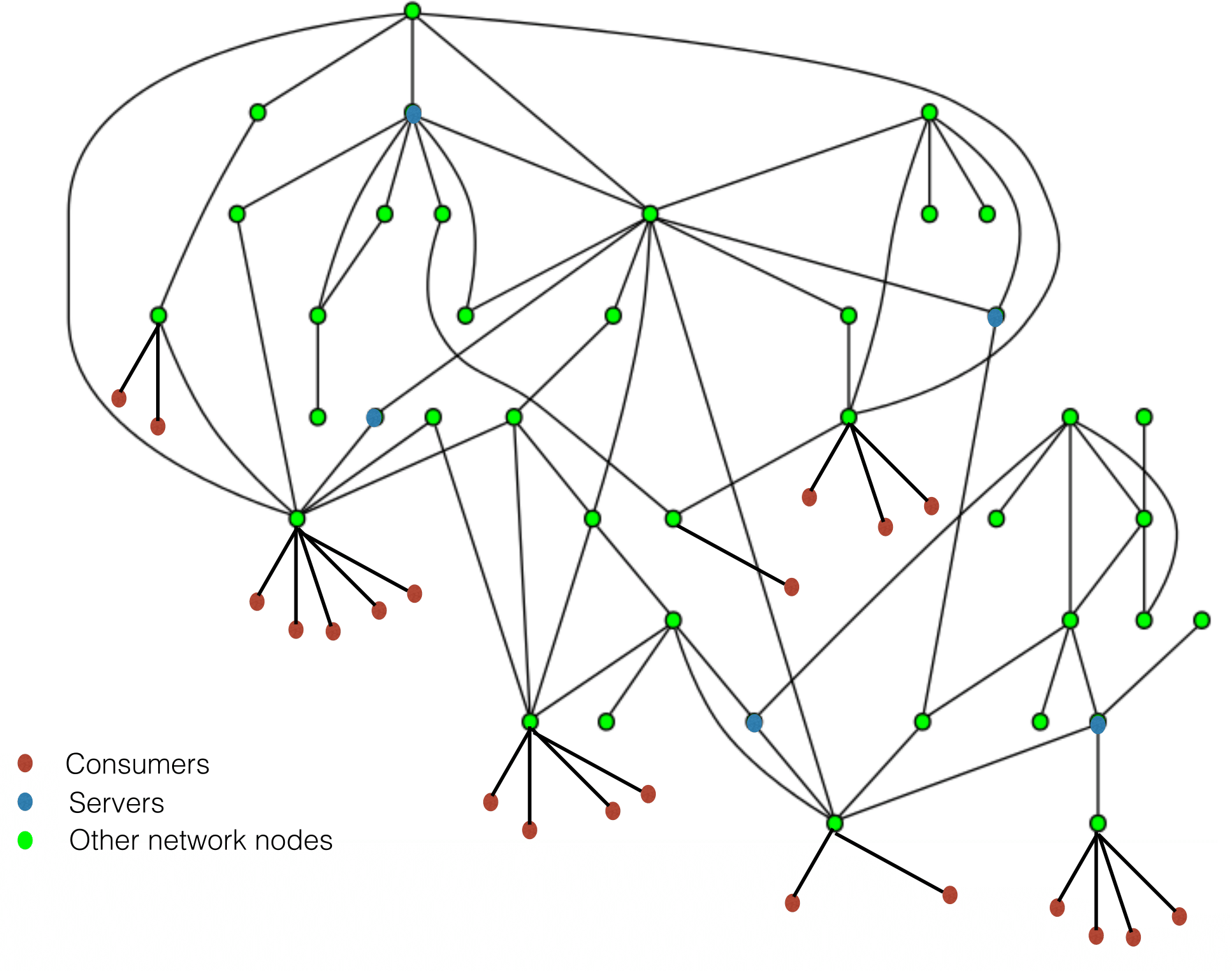}
 \caption{Geant topology and connected endpoints}
 \label{fig::geant}
  \vspace{-1em}
\end{figure}

To evaluate all the schemes, we use the GEANT network topology \cite{geant} illustrated in Fig. \ref{fig::geant}, which interconnects Europe's NRENs and provides research network services across the continent. We distribute the endpoints, {\em{i.e.}}, consumers and origin servers, randomly in each simulation. As for the consumers, we attach a variable number of nodes (between three to six nodes) to each randomly selected router. Our topology contains in total $56$ consumers. There are five origin servers, which we randomly place in the topology for each simulation. Thus, the considered topology has $101$ nodes. Therefore, our scenario is in line, in terms of type and number of nodes, with topologies used in related ICN works \cite{cobra, chiocchetti2012exploit}.
 
We test and validate all the schemes based on URLs from real traces of HTTP requests \cite{selfsimilarity}. The content universe, {\em{i.e.}}, the set of offered content objects, consists of $1000$ files. Each of the files is divided into $100$ segments. Therefore, we obtain $10^5$ unique segments in total. Each node has limited storage space and can cache up to $100$ segments in its content store. 
 In NDN, content objects are divided into segments. Since ndnSIM does not permit fragmentation, we consider the payload of each segment to be fixed. We assume that the content popularity follows the Zipf-Mandelbrot law, which is shown in $(\ref{eq::zipf})$, where $M$ denotes the cardinality of a content catalogue and is used to characterize content popularity and $\alpha$ is the skewness  of the popularity function (larger $\alpha$ values correspond to fewer popular content objects). 

\begin{equation}
\label{eq::zipf}
P(x=i) =  \frac{1/i^\alpha}{\sum_{j=1}^{M} 1/j^\alpha}
\end{equation}

The comparative analysis of BFR with flooding and shortest path routing strategies is done using $\alpha$ in the $[0.8,1.4]$ interval. All the results are averaged over ten simulations (each simulation lasts $100'000$ seconds). The reported mean values have $95\%$ confidence intervals.

 We use the BF parameter set $\{N=200, p_{fpp}=0.02\}$. $N$ denotes the inserted element count and $p_{fpp}$ denotes the false positive probability for BFs. The size of each content advertisement BF is $203.5$ bytes for advertising $200$ URLs.

\subsection{Results}
 We evaluate all the schemes based on the following performance metrics: 1) average round-trip delay, 2) robustness to topology changes, 3) communication overhead, and 4) mean hit distance. Further, we present results concerning the impact of false positive reports from BFs on BFR routing for different levels of the false positive error. In the following, we discuss results for these metrics.

\subsubsection{Average round-trip delay}
\begin{figure*}[!t]
 \centering
 \subfloat[][]{\label{fig::delay}\includegraphics[width=0.35\textwidth]{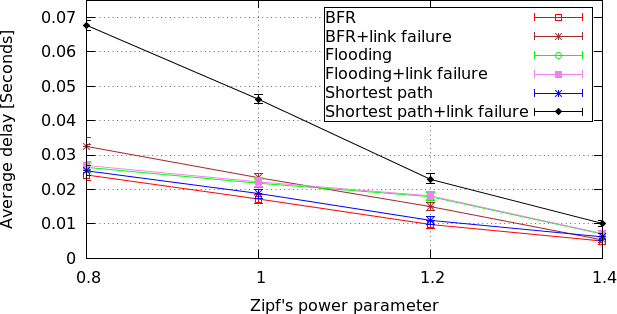}}
 \subfloat[][]{\label{fig::failureUnsatisfaction}\includegraphics[width=0.32\textwidth]{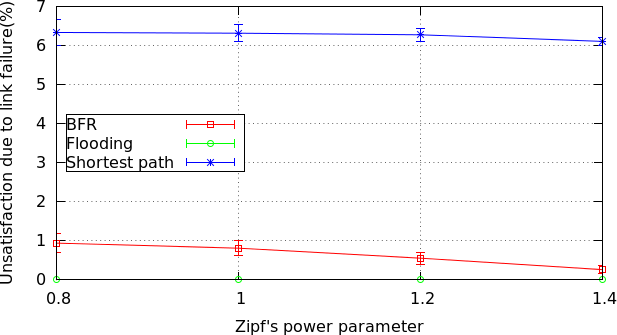}}
 \subfloat[][]{\label{fig::overhead}\includegraphics[width=0.33\textwidth]{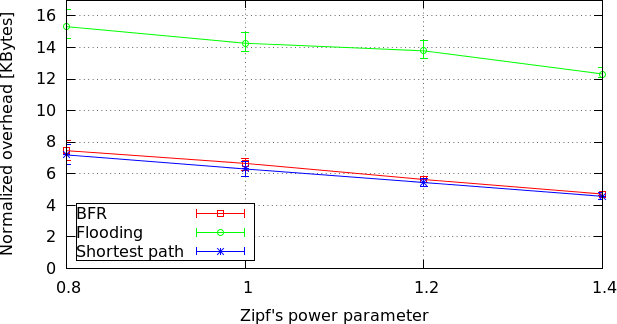}}
 \caption[]{Results for different values of $\alpha$ : (a) Average round-trip delay without and in presence of link failures ; (b)  Impact of link failure on Interest unsatisfaction; (c) Normalized communication overhead} 
 \label{fig:res}
 \vspace{-1.5em}
\end{figure*}

We evaluate the performance of all the schemes under comparison in terms of average round-trip delay, {\em{i.e.}}, the average delay from the time instant consumers send Interests until the time they retrieve the demanded content objects. To better show the behaviour of all the considered schemes in the presence of topology changes, we also measure the average round-trip delay in presence of link failures for all the schemes. We schedule three link failures at time instants $5'000$, $15'000$, and $25'000$. These links recover at time instants $10'000$, $20'000$, and $30'000$ respectively. Fig. \ref{fig::delay} illustrates the results for average round-trip delays. From this figure, we observe that flooding shows the highest delay in absence of link failures. The reason is that flooding all the Interests creates bottlenecks and results in high delays. The shortest path approach has lower average delay compared to flooding. The reason is that it forwards each Interest only through the face that has the shortest path to the origin server. This is not always efficient as the shortest path is not always the ``best" path. In \cite{jim1}, the authors show that the ``best" path is the one with the highest throughput, or the least congested path in other words. BFR benefits from multipath communication and hence forwards Interests through all the faces that the demanded content object can be reached with high probability. When the shortest paths are congested, BFR also exploits longer, but less congested paths for sending the Interests and thus performs better than shortest path routing in terms of delay.


In presence of link failures, Fig. \ref{fig::delay} confirms the resilience of flooding to the link failures because it broadcasts the Interests and forwards them over all the paths. This figure shows that BFR is also resilient to link failures in terms of delay. This is due to the fact that BFR benefits from the existence of multiple paths towards origin servers and does not forward the Interests over a single path. From Fig. \ref{fig::delay}, we can also observe that shortest path routing is the less resilient approach to link failures. This is because it always relies on a single path and forwards the Interests over this path to the origin server of the demanded content objects, while a link failure might occur on that path.
\begin{figure*}[!t]
 \centering
 \subfloat[][]{\label{fig::totalInterestOverhead}\includegraphics[width=0.33\textwidth]{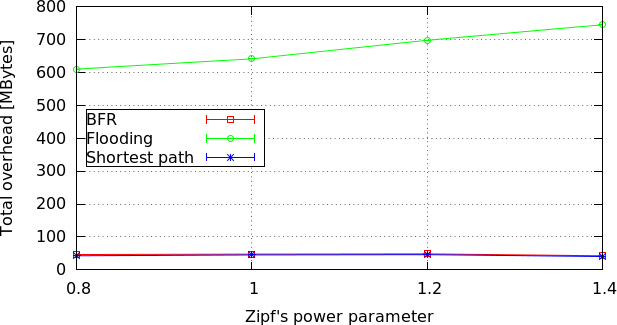}}
\subfloat[][]{\label{fig::routingOverhead}\includegraphics[width=0.33\textwidth]{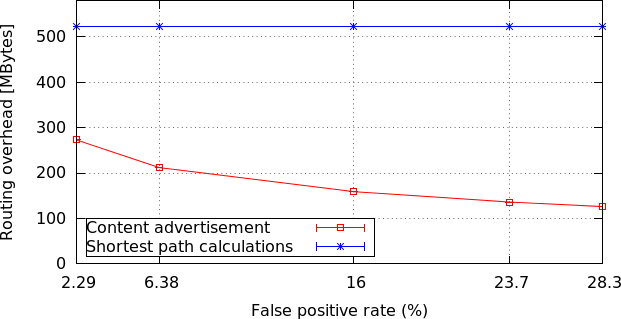}}
  \subfloat[][]{\label{fig::hitdistance}\includegraphics[width=0.33\textwidth]{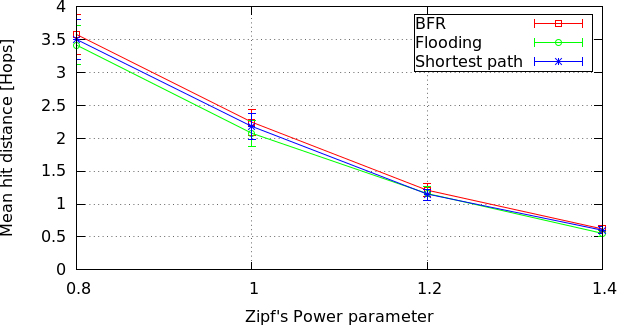}}
  \caption[]{Results for : (a) Total Interest overhead for different values of $\alpha$; (b) A comparison of content advertisement overhead for different levels of false positive errors and the needed overhead for calculating shortest paths; (c)  Mean hit distance for different values of $\alpha$}
 \label{fig:res}
\end{figure*}

\subsubsection{Robustness to topology changes}
In Fig. \ref{fig::delay}, we illustrated the impact of link failures on average round-trip delay for all the schems under comparison. In Fig. \ref{fig::failureUnsatisfaction}, we compare the performance of all the considered schemes in terms of the impact of link failures on the percentage of unsatisfied Interests for different  values of $\alpha$. Fig. \ref{fig::failureUnsatisfaction} shows that all the Interests are satisfied in presence of link failures when flooding is used because it broadcasts the Interests and does not rely only on the paths on which links have failed. Using BFR, the maximum rate of unsatisfied Interests is only $0.93\%$. This is attributed to the fact that BFR forwards Interests over all the paths en-route to the origin servers of demanded content objects. Also from Fig. \ref{fig::failureUnsatisfaction}, we can see that maximum rate of unsatisfied Interests for shortest path routing is approximately $6.4\%$. The performance of shortest path routing degrades in the presence of link failures because it always relies on the shortest path towards the origin server of the demanded content object on which links might fail.

\subsubsection{Communication overhead}
Fig. \ref{fig::overhead} illustrates results concerning the normalized communication overhead for retrieving a Data packet, {\em{i.e.}}, the summation of the total communication overhead for forwarding all the Interests and Data packets divided by the number of retrieved Data packets. From Fig.  \ref{fig::overhead}, we observe the very high communication overhead for flooding. This is due to the forwarding of each incoming Interest to all the available faces except from the incoming one. This forwarding strategy wastes an enormous amount of bandwidth and also has unnecessary storage overhead for nodes that are not situated towards the origin servers or will not receive a copy of the demanded content object in the foreseeable future. We also see from Fig. \ref{fig::overhead} that BFR and shortest path have quite close normalized communication overhead. Fig. \ref{fig::totalInterestOverhead} illustrates results in terms of total communication overhead needed for forwarding Interests for different values of $\alpha$. As Fig. \ref{fig::totalInterestOverhead} shows, BFR and shortest path need on average only $6.9\%$ and $6.5\%$ of the communication overhead that flooding requires for boradcasting Interests, respectively.



Fig. \ref{fig::routingOverhead} illustrates the total communication overhead needed for propagating content advertisements in BFR for different levels of false positive error probability as well as the required communication overhead for calculating shortest paths in the shortest path approach. For BFR, we consider four sets of parameters for content advertisement BFs as shown in Table \ref{tab:wrongRouting}. From this Table, it is evident the trade-offs between different numbers of hash functions $(k)$, different overhead values per inserted element $(m/n)$, and different values of false positive error probability. As Fig. \ref{fig::routingOverhead} shows, the communication overhead required for calculating shortest paths in shortest path routing is on average approximately three times more than the communication overhead required for propagating content advertisements in BFR.

\subsubsection{Mean hit distance}
We present results in Fig. \ref{fig::hitdistance} concerning the mean hit distance, {\em{i.e.}}, the number of hops that an Interest has to travel to reach the demanded content object. As Fig. \ref{fig::hitdistance} shows, all the considered schemes perform very close to each other in terms of mean hit distance. The flooding approach has a slightly better performance for $\alpha=0.8$ and $\alpha=1$. However, for $\alpha=1.2$ and $\alpha=1.4$, all the schemes perform approximately equal in terms of mean hit distance. This is due to the fact that when the value of $\alpha$ grows, a smaller set of content objects are popular that most of them are cached close to the consumers.

\begin {table}[!t]
\caption {False positive error probability under various $m/n$ and $k$ combinations} 
\label{tab:wrongRouting} 
\begin{center}
\begin{tabular}{l*{6}{c}r}
$m/n$ & $k$ & $p_{fpp}$  \\ \hline
$3$ & $2$ & $28.3\%$  \\ 
$3$ & $2$ & $23.7\%$ \\
$4$ & $3$ & $16.0\%$ \\
$6$ & $4$ & $6.38\%$ \\
 $8$ & $5$ & $2.29\%$ \\ \hline
\end{tabular}
\end{center}
  \vspace{-1.5em}
\end {table}

The shortest path routing scheme using the Dijkstra algorithm requires accurate information regarding the topology of the network to determine the shortest paths. On the other hand, BFR operates without having any information about the network topology. Further, the shortest path scheme uses the {\emph{best path}} forwarding strategy. This strategy requires routing protocols to calculate the shortest paths towards origin servers. One possibility is to employ an IP-based routing protocol for such a purpose as a fall-back or main mechanism to work in parallel with name-based routing. Employing classical IP-based routing protocols in NDN entails scalability issues \cite{nlsr} and imposes significant signalling overhead even for intra-domain scenarios. This is another advantage of BFR compared to the shortest path approach.

 \subsubsection{Impact of false positive errors on BFR operation}
We present results concering the impact of false positive errors on BFR operation in terms of percentage of Interests that have been routed towards both correct and wrong origin servers for different values of false positive error probabilities shown in Table  \ref{tab:wrongRouting}.
 \begin{figure}[t]
  \vspace{-0.5em}
\centering
 \includegraphics[width=0.66\columnwidth]{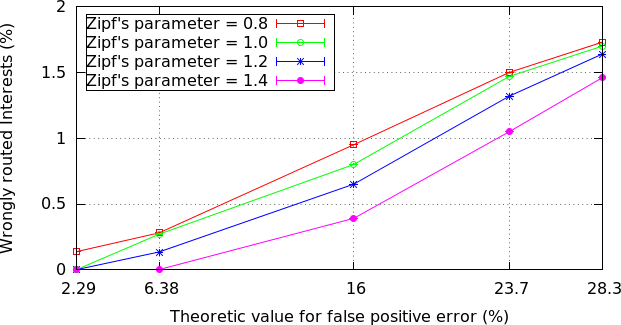}
 \caption{Impact of false positive reports on BFR routing for different values of $p_{fpp}$ and $\alpha$}
 \label{fig::wrongRouting}
  \vspace{-1em}
\end{figure} 

Fig. \ref{fig::wrongRouting} shows that the higher the probability of false positive error is, the higher are the number of Interests that not only have been routed towards correct origin servers, but have reached wrong origin servers as well. Further, Fig. \ref{fig::wrongRouting} shows the impact of increasing the value of $\alpha$ on the percentage of Interests that are also routed towards wrong origin servers. We note that when the value of $\alpha$ is higher,  a smaller set of content objects are popular and this results in measuring less false positive reports in practice. We observe the highest impact of false positive reports on BFR routing for $p_{fpp}=28.3\%$ and $\alpha=0.8$, when only $1.73\%$ of Interests are routed also towards wrong origin servers. Note that all the Interests are satisfied in the presence of false positive reports and the only practical impact of false positive reports is that a very small number of Interests reach wrong origin servers, {\em{i.e.}}, the origin servers that do not provide the demanded content objects, while all the Interests are routed towards correct origin servers, {\em{i.e.}}, the origin servers that provide the demanded content objects.

\section{Conclusion}
\label{sec::con}
In this work we proposed BFR, a BF-based, fully distributed, content oriented, and topology agnostic routing approach at the intra-domain level for NDN. Our approach is based on propagation of content advertisements from origin servers using BFs. BFR incurs small storage overhead as well as reasonable signalling overhead. BFR outperforms flooding and shortest path approaches in terms of  communication cost and average round-trip delay. In terms of  robustness to topology changes, our scheme strongly outperforms the shortest path approach. BFR does not require any auxiliary routing protocols for calculating best paths, which is in contrast to schemes based on the shortest path. Our scheme does not adopt IP-based routing protocols as a primary or fall-back mechanism. 

Our future work includes desgining storage management strategies for CAI messages based on BF aggregation, especially when the content universe size increases. Further, we aim at examining BFR with other forwarding strategies and compare its performance with other NDN routing protocols. We also intend to test and evaluate our scheme using other realistic content catalogues, topologies, and scenarios as described in \cite{icnirtf}.







\bibliographystyle{IEEEtran}
\bibliography{myrefs.bib}

\begin{thebibliography}{10}
\providecommand{\url}[1]{#1}
\csname url@samestyle\endcsname
\providecommand{\newblock}{\relax}
\providecommand{\bibinfo}[2]{#2}
\providecommand{\BIBentrySTDinterwordspacing}{\spaceskip=0pt\relax}
\providecommand{\BIBentryALTinterwordstretchfactor}{4}
\providecommand{\BIBentryALTinterwordspacing}{\spaceskip=\fontdimen2\font plus
\BIBentryALTinterwordstretchfactor\fontdimen3\font minus
  \fontdimen4\font\relax}
\providecommand{\BIBforeignlanguage}[2]{{%
\expandafter\ifx\csname l@#1\endcsname\relax
\typeout{** WARNING: IEEEtran.bst: No hyphenation pattern has been}%
\typeout{** loaded for the language `#1'. Using the pattern for}%
\typeout{** the default language instead.}%
\else
\language=\csname l@#1\endcsname
\fi
#2}}
\providecommand{\BIBdecl}{\relax}
\BIBdecl

\bibitem{ICNsurvey}
B.~Ahlgren, C.~Dannewitz, C.~Imbrenda, D.~Kutscher, and B.~Ohlman, ``A survey
  of information-centric networking,'' \emph{IEEE Communications Magazine},
  vol.~50, no.~7, pp. 26--36, Jul. 2012.

\bibitem{locationindependence}
Z.~Gao, A.~Venkataramani, J.~F. Kurose, and S.~Heimlicher, ``Towards a
  quantitative comparison of location-independent network architectures,''
  \emph{ACM SIGCOMM Computer Communication Review}, vol.~44, no.~4, pp.
  259--270, Oct. 2015.

\bibitem{nlsr}
A.~Hoque, S.~O. Amin, A.~Alyyan, B.~Zhang, L.~Zhang, and L.~Wang, ``{NLSR}:
  named-data link state routing protocol,'' in \emph{Proc. of the 3rd ACM
  SIGCOMM workshop on Information-centric networking}, Aug. 2013, pp. 15--20.

\bibitem{advertising}
Y.~Wang, K.~Lee, B.~Venkataraman, R.~L. Shamanna, I.~Rhee, and S.~Yang,
  ``Advertising cached contents in the control plane: Necessity and
  feasibility,'' in \emph{IEEE Conference on Computer Communications Workshops
  (INFOCOM WKSHPS)}, Mar. 2012, pp. 286--291.

\bibitem{scan}
M.~Lee, K.~Cho, K.~Park, T.~T. Kwon, and Y.~Choi, ``{SCAN}: Scalable content
  routing for content-aware networking,'' in \emph{Proc. of the IEEE Int. Conf.
  on Communications (ICC)}, Jun. 2011, pp. 1--5.

\bibitem{cobra}
M.~Tortelli, L.~A. Grieco, G.~Boggia, and K.~Pentikousisy, ``{COBRA}: Lean
  intra-domain routing in ndn,'' in \emph{Proc. of the IEEE 11th Consumer
  Communications and Networking Conference (CCNC)}, Jan. 2014, pp. 839--844.

\bibitem{netinf}
C.~Dannewitz, D.~Kutscher, B.~Ohlman, S.~Farrell, B.~Ahlgren, and H.~Karl,
  ``Network of information ({N}et{I}nf)--an information-centric networking
  architecture,'' \emph{Computer Communications}, vol.~36, no.~7, pp. 721--735,
  Apr. 2013.

\bibitem{mdht}
M.~D'Ambrosio, C.~Dannewitz, H.~Karl, and V.~Vercellone, ``{MDHT}: a
  hierarchical name resolution service for information-centric networks,'' in
  \emph{Proc. of the ACM SIGCOMM workshop on Information-centric networking},
  Aug. 2011, pp. 7--12.

\bibitem{pursuit}
D.~Trossen, G.~Parisis, K.~Visala, B.~Gajic, J.~Riihijarvi, P.~Flegkas,
  P.~Sarolahti, P.~Jokela, X.~Vasilakos, C.~Tsilopoulos \emph{et~al.},
  ``Pursuit conceptual architecture: pinciples, patterns and sub-components
  descriptions,'' Tech. Rep., May 2011.

\bibitem{breadcrumbs}
E.~J. Rosensweig and J.~Kurose, ``Breadcrumbs: Efficient, best-effort content
  location in cache networks,'' in \emph{Proc. of the IEEE INFOCOM}, 2009, pp.
  2631--2635.

\bibitem{inform}
R.~Chiocchetti, D.~Perino, G.~Carofiglio, D.~Rossi, and G.~Rossini, ``Inform: a
  dynamic interest forwarding mechanism for information centric networking,''
  in \emph{Proc. of the ACM 3rd SIGCOMM workshop on Information-centric
  networking}, Aug. 2013, pp. 9--14.

\bibitem{dht}
H.~Liu, X.~De~Foy, and D.~Zhang, ``A multi-level {DHT} routing framework with
  aggregation,'' in \emph{Proc. of the ACM 2nd edition of the ICN workshop on
  Information-centric networking}, Aug 2012, pp. 43--48.

\bibitem{scopedflooding}
L.~Wang, S.~Bayhan, J.~Ott, J.~Kangasharju, A.~Sathiaseelan, and J.~Crowcroft,
  ``Pro-diluvian: Understanding scoped-flooding for content discovery in
  information-centric networking,'' in \emph{Proc. of the ACM 2nd Int. Conf. on
  Information-Centric Networking}, Sep. 2015, pp. 9--18.

\bibitem{prefetching}
A.~W. Kazi, ``Prefetching bloom filters to control flooding in content-centric
  networks,'' in \emph{Proc. of the ACM CoNEXT Student Workshop}, Nov. 2010,
  p.~22.

\bibitem{summarycache}
L.~Fan, P.~Cao, J.~Almeida, and A.~Z. Broder, ``Summary cache: a scalable
  wide-area web cache sharing protocol,'' \emph{IEEE/ACM Transactions on
  Networking (TON)}, vol.~8, no.~3, pp. 281--293, Jun. 2000.

\bibitem{bloom}
B.~H. Bloom, ``Space/time trade-offs in hash coding with allowable errors,''
  \emph{Communications of the ACM}, vol.~13, no.~7, pp. 422--426, Jul. 1970.

\bibitem{selfsimilarity}
M.~E. Crovella and A.~Bestavros, ``Self-similarity in world wide web traffic:
  evidence and possible causes,'' \emph{IEEE/ACM Transactions on Networking
  (TON)}, vol.~5, no.~6, pp. 835--846, Dec. 1997.

\bibitem{stateful}
C.~Yi, A.~Afanasyev, I.~Moiseenko, L.~Wang, B.~Zhang, and L.~Zhang, ``A case
  for stateful forwarding plane,'' \emph{Computer Communications}, vol.~36,
  no.~7, pp. 779--791, Apr. 2013.

\bibitem{ndnsim2.1}
S.~Mastorakis, A.~Afanasyev, I.~Moiseenko, and L.~Zhang, ``ndn{SIM} 2.0: A new
  version of the {NDN} simulator for {NS}-3,'' Tech. Rep., Jan. 2015.

\bibitem{geant}
``The geant network, 2012,'' \url{http://www.topology-zoo.org/dataset.html},
  accessed: 2016-07-25.

\bibitem{chiocchetti2012exploit}
R.~Chiocchetti, D.~Rossi, G.~Rossini, G.~Carofiglio, and D.~Perino, ``Exploit
  the known or explore the unknown?: {H}amlet-like doubts in {ICN},'' in
  \emph{Proc. of the ACM 2nd edition of the ICN workshop on Information-centric
  networking}, Aug. 2012, pp. 7--12.

\bibitem{jim1}
M.~Badov, A.~Seetharam, J.~Kurose, V.~Firoiu, and S.~Nanda, ``Congestion-aware
  caching and search in information-centric networks,'' in \emph{Proc. of the
  ACM 1st international conference on Information-centric networking}, Sep.
  2014, pp. 37--46.

\bibitem{icnirtf}
``Information-centric networking: Baseline scenarios
  draft-irtf-icnrg-scenarios-01,''
  \url{https://www.ietf.org/proceedings/88/id/draft-irtf-icnrg-scenarios-01.txt},
  2016-07-16.

\end{thebibliography}
%
%
%

\end{document}